# Experimental procedures for precision measurements of the Casimir force with an Atomic Force Microscope

**Hsiang-Chih Chiu, Chia-Cheng Chang, R Castillo-Garza, F Chen, and U Mohideen**
Department of Physics and Astronomy, University of California, Riverside, CA 92521.

Email:umar.mohideen@ucr.edu

**Abstract.** Experimental methods and procedures required for precision measurements of the Casimir force are presented. In particular, the best practices for obtaining stable cantilevers, calibration of the cantilever, correction of thermal and mechanical drift, measuring the contact separation and the roughness are discussed.

## 1. Introduction

The role of the quantum vacuum and its modification by boundaries, generally referred to as the Casimir effect [1-6] is finding ever increasing applications in fields extending from cosmology to nanotechnology. In fundamental physics the Casimir force has been used for setting limits on the existence of extra dimensions and forces outside the standard model [7-15]. As the Casimir force dominates the interaction at separation distances less than 100 nm, its precision measurement is an effective probe of new physics at short distance scales. An interesting feature of the Casimir force is its strong material and geometry dependence. Theoretically the Casimir force can even be made repulsive through a judicious choice of materials [2,16] or boundary shape [3-6]. The significant role of the Casimir force has also been realized in nanotechnology, where operating surfaces in micromachines are separated by distances less than 1 micron [17,18]. Given the trend towards decreasing dimensions in nanotechnology, it is very conceivable that in the near future the distance scales will drop below 100nm. In this case, the Casimir force will have an overwhelming influence on both the design and function of microelectromechanical systems (MEMS).

An extensive review of older experiments is provided in Ref. [6]. Precision measurements in vacuum over the last five years have used the Atomic Force Microscope (AFM) [19-24] and MEMS [25,26]. Given, the general importance and the novel applications of the Casimir force, many new precision experiments are planned [27,28]. On a more immediate level these experiments are expected to check the more subtle geometry and material dependences of the Casimir force. In the latter case, the most pressing issue is the deviation from the proximity force approximation for the configuration of a sphere and plate, that is most often used in experiments. Much theoretical effort has been expended in the last two years in advancing more precise methods of calculating the Casimir force between a sphere and a plate[29-31]. Unambiguous observation of these effects would require precision better than 0.1% for the short distances at which AFM and MEMS devices have been applied [19-22, 32-35]. Another important experimental configuration of the Casimir force between a cylinder and plate is exactly solvable [30,36,37]. However, this configuration is experimentally challenging as it requires the perfect alignment of the cylinder axis with the surface of the plate [27]. Some relevant results were obtained also using a semiclassical geometric optic approach [38]. The material dependence of the Casimir force is more important from a technological point of view [19-22]. While using dielectric boundaries, would lead to uncontrollable electrostatic forces which cannot be easily compensated, even experiments using, non-metallic but conductive materials such as semiconductors require special precautions [19-22]. Improvements in many aspects of the measurement are necessary to achieve the next stage.

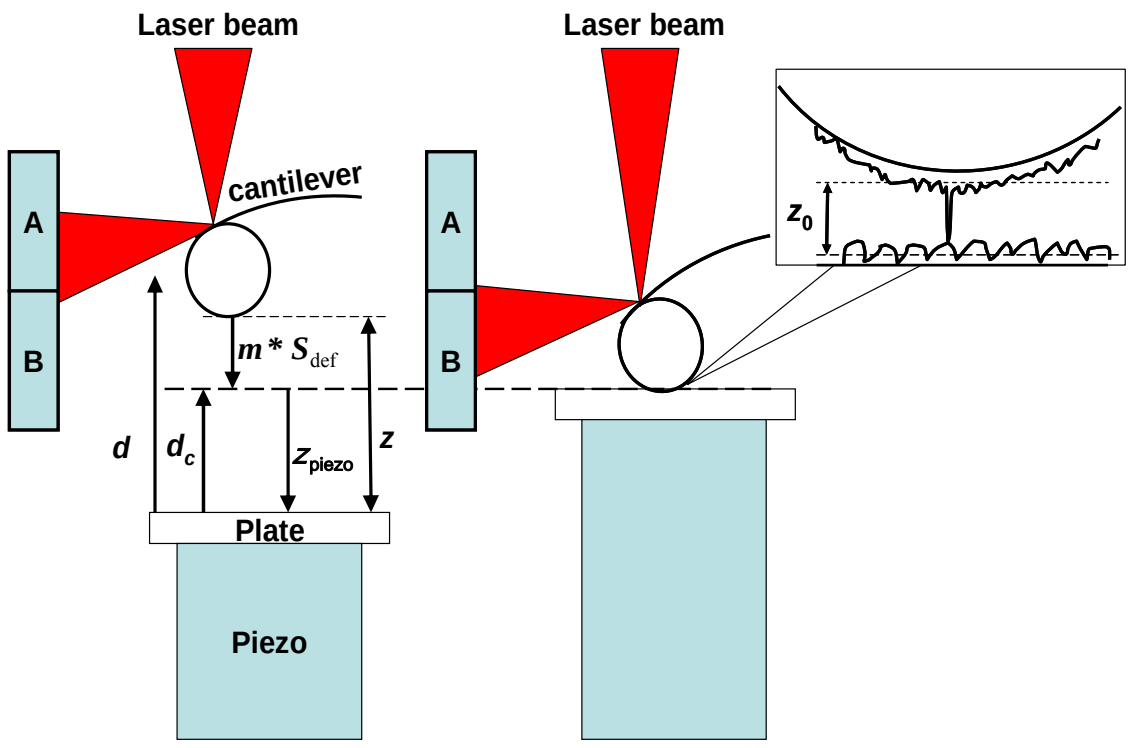

Figure 1: Schematic of setup used. The extension of the piezo is given by $d$, with $d_c$ being its value at the point of sphere-plate contact, $z_{piezo}$ is the distance moved by the plate measured from contact, $z_0$ is the average separation on contact, $m$ is the deflection coefficient and $S_{def}$ is the deflection signal.

The AFM is one of the most important tools of choice for precision measurements of the Casimir force primarily due to its exquisite force detection sensitivity of around $10^{-18}$ N [39]. A schematic (not to scale) is shown in figure 1. The force detection is based on measuring the deflection of a micron thick microfabricated cantilever. In the cases discussed here, the deflection is measured by monitoring the deviation of a laser beam reflected off the top of the cantilever with two photodiodes A and B. This deflection signal $S_{def}$ is calibrated using electrostatic means described here. The Casimir force is measured as a function of the sphere plate separation $z$, by moving the plate with the piezo closer to the sphere till they make contact. An important advantage of the AFM technique is that, if this contact point $d_c$, can be precisely determined, it can be used as a reference in averaging many experimental repetitions leading to substantial reductions in the random noise.

All aspects of AFM precision Casimir force measurements, from sample preparation to the measurement procedure are discussed below. Particular emphasis is given to the calibration, identification and removal of systematic errors due to residual potential differences, thermal and mechanical drift. The vacuum requirements are discussed in section 2. Section 3 presents the procedure followed for the making the metal coated cantilever-sphere system. In section 4 the calibration of the distance $d$ moved by the piezo used for changing the sphere plate separation is discussed. Section 5 presents the use of the electrostatic force measurement between the sphere and plate, used for calibrating the cantilevers, measuring the average surface separation on contact of the two surfaces $z_0$ and obtaining the value of the residual potential difference between the two surfaces. It can also be used to cross-check the value of the Casimir force. Section 6 discusses the surface roughness measurement. Section 7 is a very brief discussion of the Casimir force measurement, which has been extensively discussed in many of our recent publications [19-22,40]. The measurement of the lateral Casimir force is not discussed here [41,42]. Section 8 is the conclusion.

## 2. **Vacuum Requirements**

It is important that the boundary surfaces used in Casimir force measurements be preserved in the cleanest state possible, as the optical properties of the material are directly input into the calculation of the Casimir force. The purity of the sample surface has to be preserved for periods of even a few days, as the need to reach a stable thermal and mechanical equilibrium necessitates extended experimental time scales. This requires the use of low pressures and a high vacuum in the experimental chamber. The experimental chamber should be of nonmagnetic stainless steel capable of being baked to 125°C to rid it of adsorbed volatiles on the walls. The AFM to be used should be constructed or modified to be free of volatile organics and any elements that might outgas during the pumpdown or experiment. Typically, only the laser and optical detection elements of the AFM should be present in the vacuum. A typical vacuum setup used in our lab for precision measurements is shown below.

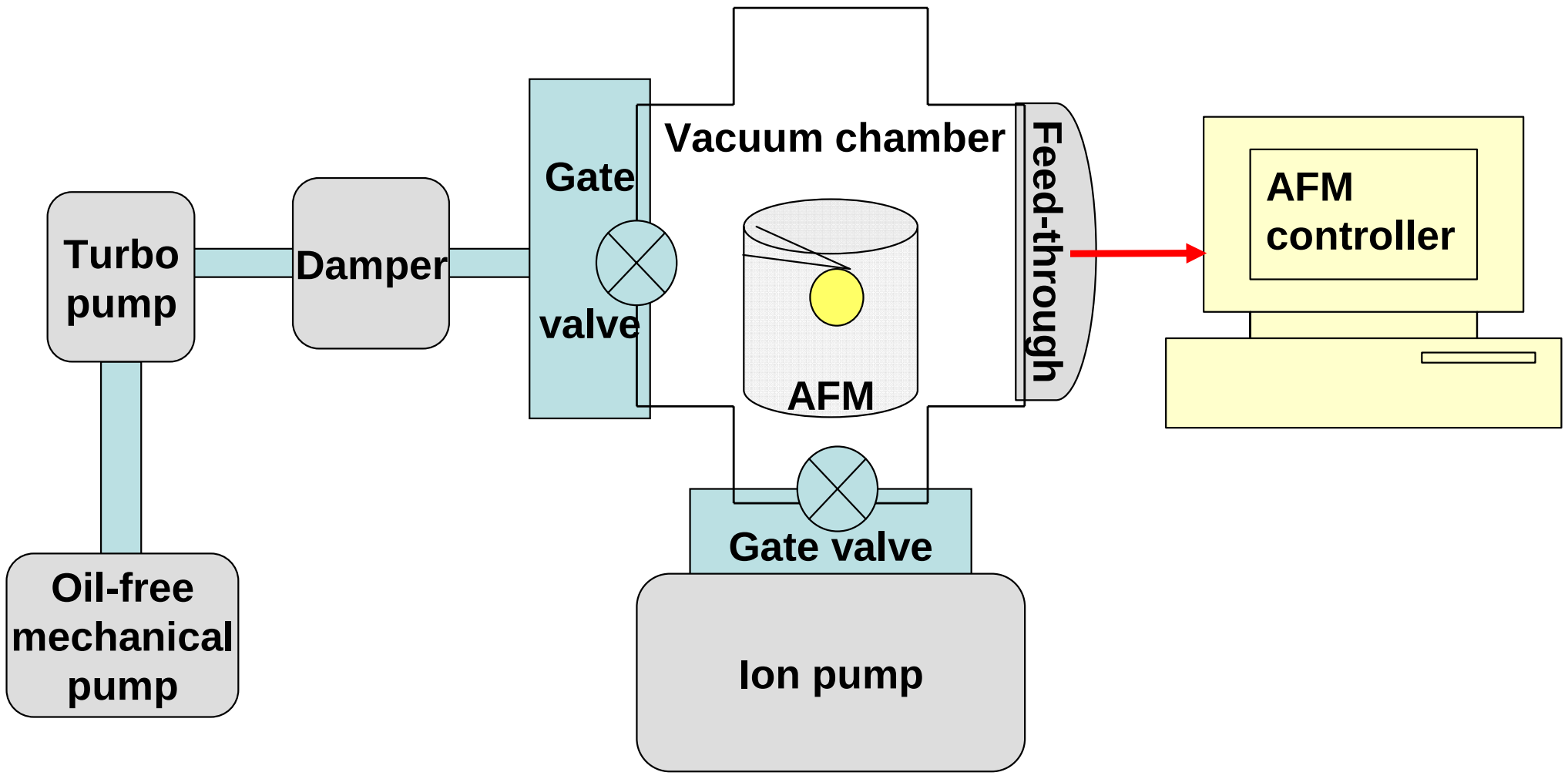

Figure 2: Layout of the vacuum setup used in precision measurements of the Casimir force.

A pressure of at least $10^{-7}$ Torr is necessary for preserving the sample quality for a 2-week period. While pressures of $10^{-11}$ Torr are feasible, it is not necessary for the typical experimental durations necessary in this case. The most important detail to notice in figure 2 is that it is an "oil-free" system. No diffusion pumps or standard mechanical pumps are used. Only turbo mechanical pumps, ion pumps and oil-free mechanical pumps are acceptable. The experimental chamber should be supported on a passively damped optical table to reduce the mechanical noise. The turbomechanical pump and the mechanical pump lead to the introduction of mechanical noise into the AFM system. Therefore during data acquisition, only the ion pump should be used and the turbo pump needs to be valved and turned off.

### 3. Preparation of the Au coated sphere-cantilever system

Uncoated triangular silicon nitride cantilevers greater than 300 μm long are preferred as they possess the lowest spring constants and have excellent lateral stability. First a 30 nm layer of Al is coated on the cantilever. Thermal evaporation is used for the Al coating and the cantilevers are rotated during the coating process to ensure uniformity. Next, in order to attach the sphere, about 20 x 20 x 20 $\mu m^3$ spot of conductive silver epoxy is applied to the tip of the cantilever under an optical microscope. Excessive use of epoxy should be avoided and special caution should be used to prevent the epoxy from spreading to the other side of the cantilever. Using a stripped optical fiber, a 200 μm diameter polystyrene sphere is picked up and placed on the tip of the cantilever. Such polystyrene spheres are ideal as they are light weight, have a smooth surface and have very little eccentricity (less than $10^{-3}$). As the Casimir force increases with the sphere diameter, we have experimented with larger polystyrene spheres, even as large as 1mm. However, the larger spheres have a higher surface roughness and are not appropriate for precision measurements. Even 200 μm diameter spheres from different suppliers or different batches from the same supplier exhibit a large degree of variability in their surface roughness. We have found that the 200 μm spheres from Duke Scientific, CA, made before 1999 to have the smoothest surfaces. After attachment of the sphere, the epoxy is allowed to cure in a clean environment. Next, the whole cantilever-sphere system is coated with > 80nm layer of Au in a thermal evaporator. Uniformity of Au coating is ensured through rotation. In the AFM (figure 1), the laser is reflected off the top of the cantilever and the resultant asymmetric heating and thermal expansion leads to cantilever deformation. This in turn leads to thermal and mechanical instabilities which are exacerbated in high vacuum. We have found that this combination of Al and Au on the cantilever will make the deflection signals stable in vacuum. Both thermal and e-beam evaporation were tried in the past and found to be equally good. Here again oil-free vacuum chambers are used and all metal coatings are done at pressures lower than or around $10^{-6}$ Torr. Thus the vacuum chamber needs to be frequently cleaned with HCl, trichloroethylene, acetone and methanol to remove metallic and organic contaminants. We have found a definite correlation between low sphere-plate residual potential differences and the cleaning process. A ~ 0 mV residual potential differences between the Au coated sphere and plate can be achieved immediately after a cleaning procedure. In the case when Au plates are used, a 10 nm layer of Cr is applied either through thermal evaporation or sputtering prior to thermal evaporation of a 100 nm thick Au coating. Typically, highly polished sapphire plates are used as the substrate.

## 4. Piezo actuator calibration

Tube piezos capable of large extensions of order 6 μm are preferred. Such large piezo extensions provide a sufficient separation range allowing precise analysis of the long range electrostatic forces. The large piezo extensions *d*, are also necessary to allow time for the damping of the oscillations associated with the contact and separation of the sphere and plate. The separation between the bottom of the gold sphere and the top of the plate is given by:

$$z = z_{piezo} + S_{def} * m + z_o \quad , \tag{1}$$

where $z_{piezo}$ is the change in separation distance due to the movement of the plate by the piezo, $S_{def}$ is the cantilever deflection signal measured with the photodiodes A,B in figure 1, $m$ is the cantilever deflection coefficient and $z_0$ is the average separation on contact of the two surfaces. The deflection coefficient $m$ measures the bending of the cantilever due to the sphere-plate force per unit deflection signal. Attractive forces correspond to negative values of the cantilever deflection signal $S_{def}$. Thus the second term on the right of the equality in (1) measures the decrease in separation distance between the sphere bottom and plate due to the bending of the cantilever. Note $z_0 \neq 0$ due to the roughness of the metal coating on the sphere and plate. Please see figure 1 for the schematic details.

The complete extension of the piezo, *d*, must be calibrated using an optical interferometer after each experiment to obtain the best precision. The interferometer used has been described elsewhere [43]. Here the interference is obtained between the cleaved, uncoated end of the optical fiber and a low reflectivity glass plate placed on top of the piezo tube. A He-Ne laser is used as the optical source as it has precise wavelength of 632.8 nm. In an improvement over our previous work in [43], the nonlinear expansion of the piezo in response to the applied voltage is calibrated up to the 4th order term in the voltage. The interference signal $I$ measured as a function of the piezo applied voltage $V_p$ is fit to the expression:.

$$I = I_0 + \tfrac{1}{2} I_1 (1 - \gamma V_p)\{1 + \cos[\,4\pi (K_v V_p + \delta)/632.8]\} \qquad (2)$$

$$K_v = K_0 + K_1 V_p + K_2 V_p^{\,2} + K_3 V_p^{\,3} + K_4 V_p^{\,4}$$

where $\delta$ is the initial distance between the fiber end and the top surface of the glass plate at zero voltage and $\gamma$ is a correction factor to the amplitude $I_1$ of the interference signal which takes into account the small deviations from perpendicularity of the fiber to the glass plate, and the diffraction effects associated with the emission of the light from the fiber. $K_0$ is the expansion coefficient at zero volts and $K_1$, $K_2$, $K_3$, $K_4$ are the first, second, third, and fourth nonlinear terms in the applied voltage. The voltage is applied to the piezo in the form of a triangular wave of the same frequency and voltage values as used in the force measurements. The expansion constants $K_1$, $K_2$, $K_3$, $K_4$ of the piezo vary for different frequencies, maximum and minimum voltages applied to the piezo. The calibration constants are also different for positive and negative applied voltages and thus each section of the applied voltage should be fit separately and the corresponding piezo expansion constants obtained. This procedure will lead to subnanometer precision. Note that picometer precision in calibration is possible with similar interferometric techniques [44].

## 5. Electrostatic Force Measurements

The measurement of electrostatic force between the sphere and plate surfaces is central to all precision Casimir force measurements. In the case of AFM measurements it is used to:

1. Measure the residual potential difference between the sphere and plate when they are both grounded;
2. Calibrate the cantilever;
3. Measure the average separation on contact of the two surfaces.

### *5.1 Electrical connections and requirements*

As is standard practice in precision measurements, it is important to avoid or minimize any electrical grounding problems and ground loops in the whole system. Care should be taken to eliminate all Schottky barriers and confirm that all contacts are Ohmic in nature. The gold coated cantilever and sphere are rigidly attached to the grounded AFM frame with metal clips ensuring good electrical contact. The AFM has a well defined ground. All other instrumental voltages are applied with respect to the AFM's ground. To minimize ground loops, all the ground wires must come from the AFM's ground. In our case most of these wires are physically attached to the microscope. In AFM experiments [19-24,32-34] the cantilever-sphere system is grounded just by the attachment of the cantilevers to their holders. Voltages are applied to the plate. A resistor of about 1 kΩ should be connected in series with the plate. This prevents a surge in the current when the plate and sphere make contact during measurement. Otherwise, the resultant joule heating will damage the gold coating on the sphere. For connecting the metal plate, the connections can be made through gold-wire bonding or using Ag-epoxy. In the case of semiconductor plates, the wire is attached to a gold pad evaporated on the bottom surface of the plate, in order to avoid contamination of the surface passivation. The sample plate is electrically insulated from the AFM with a layer of insulating material such as Vespel. Note that the high conductivity requirements preclude the use of dielectric materials.

For the electrostatic force measurement, the sphere is kept grounded and constant voltages should be applied to the plate. The power supply used for supplying voltages to the plate should have a noise amplitude of less than ±1 μV for typical voltages of order ±1 V or less applied to the plate. The deflection of the cantilever due to the electrostatic force is measured as the plate is moved towards the sphere by the extension of the piezo. The deflection signal is measured from a separation around 6 μm to past the point of contact. In our experiments [19-21], a 0.02 Hz continuous triangular voltage signal is applied to the piezo actuator to change the sphere-plate separation. The deflection signal is acquired at 32768 equal time intervals (the highest rate possible). The measurement should be repeated around 20-30 times, each with a different constant voltage applied to the plate.

An appropriate range of dc voltages, in equal measure greater and lesser than the residual potential difference $V_0$ between the sphere and plate should be used. This range should be within ±0.5 V of $V_0$ to avoid nonlinearities in the cantilever deflection. This requires knowledge of $V_0$ which is determined by preliminary measurements of the electrostatic force as a function of dc voltage applied to the plate at some fixed sphere-plate separation. The applied voltage corresponding to the smallest electrostatic force (when the residual potential is cancelled) is the approximate value of $V_0$. A more precise determination of $V_0$ used for applying compensating voltages in the Casimir force measurement is described below.

*5.2 Measurement of the residual potential difference $V_0$ between sphere and plate*

From the electrostatic force measurements, the residual potential difference between the grounded sphere and plate, $V_0$, the cantilever deflection coefficient *m*, and the cantilever spring constant *k*, are found by fitting the cantilever deflection signal,

$$S_{def} = S_0 + \frac{F_e}{km} \qquad (3)$$

where $S_0$ is the plate voltage independent offset which should be equal to the Casimir force and $F_e$ is the electrostatic force corresponding to an applied constant voltage *V* applied to the plates, given by:

$$F_{elec} = X(z)\left(V - V_o\right)^2 \qquad (4)$$

$$X(z) = -2\pi\varepsilon_0 \sum_{m=0}^{7} A_m t^{m-1}$$

where $t = \left(\frac{z}{R}\right)$, *R* is the sphere radius measured with an electron microscope (section 8) and the coefficients $A_0$ through $A_7$ are given by 0.5, -1.18260, 22.2375, -571.366, 9592.45, -90200.5, 383084., -300357 respectively. This perturbative expansion when compared to the complete sphere-plate electrostatic force expression has a relative error of $4.7 \times 10^{-5}$ and $1.5 \times 10^{-5}$ at separation distances of 1.5 μm and 5.0 μm respectively.

The parabolic dependence of the signal on the applied voltage *V* in (4) is used to obtain the residual potential difference $V_0$ between the grounded sphere and the plate. For each value of the separation distance *z*, the deflection signal $S_{def}$ is plotted as a function the applied plate voltage as shown in figure 3 . A least $X^2$ fitting procedure is then used to obtain the voltage value at the maximum in the parabola, $V_0$ and *X(z)*. The values obtained for the particular set of data in figure 3 are shown. The numbers on the data points indicate the time sequence of the applied voltages to the plate. The parabola is repeated at every *z* and the $V_0$ is measured as a function of *z*. The average value of $V_0$ so determined is the residual potential difference. Note that at this point in the analysis the exact value of *z* is uncertain as the average separation on contact $z_0$ has not yet been determined. A plot of $V_0$ as a function of the separation distance *z* is shown in figure 4 for the case of the semiconductor plate of conductivity 3.2 x $10^{20}$ $cm^{-3}$ and a gold coated sphere both used in one of our recent measurements [22]. The larger random error with increasing separation is due to the decrease in the signal to noise ratio. In this experiment the average $V_0$ was determined to be -0.337±0.002 V. An important feature to be noticed in figure 4 is the relative constant average value of $V_0$ as a function of the separation. The value changes only within the resolution error. This is a basic and necessary condition for every Casimir force measurement. If the $V_0$ is not independent of separation it indicates the presence of electrostatic surface impurities, space charge effects [45] and/or electrostatic inhomogeneities on the sphere or plate surface.

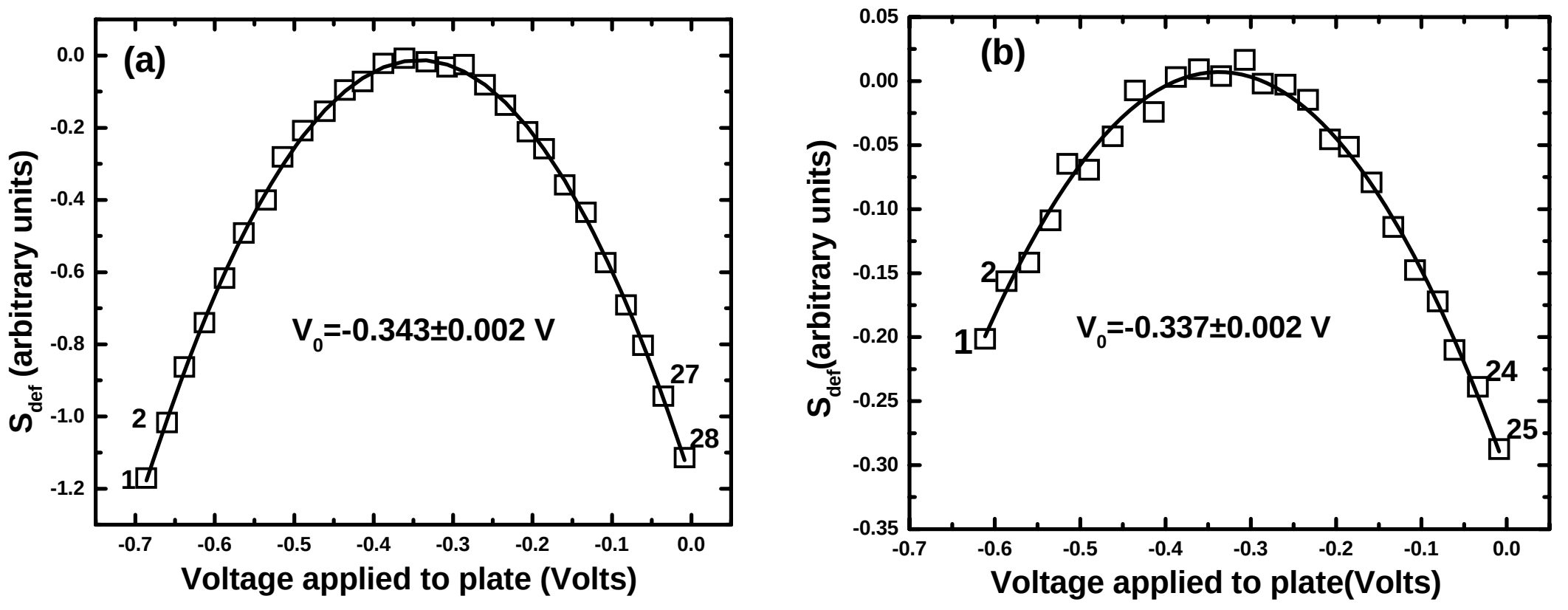

Figure 3: The deflection signal plotted as a function of the plate applied voltage at a fixed separation distance $z$ for a silicon plate with resistivity (a) ρ=0.43 Ω cm, (b) ρ=6.7 x $10^{-4}$ Ω cm used in Ref. [22]. The gold coated sphere was grounded. The applied constant voltages ranged between -0.712 to -0.008 V in (a) and -0.611 to -0.008 V in (b). The $V_o$ values are obtained are shown for this particular fit. The numbers on the data points correspond to the time sequence of the applied constant voltages to the plate.

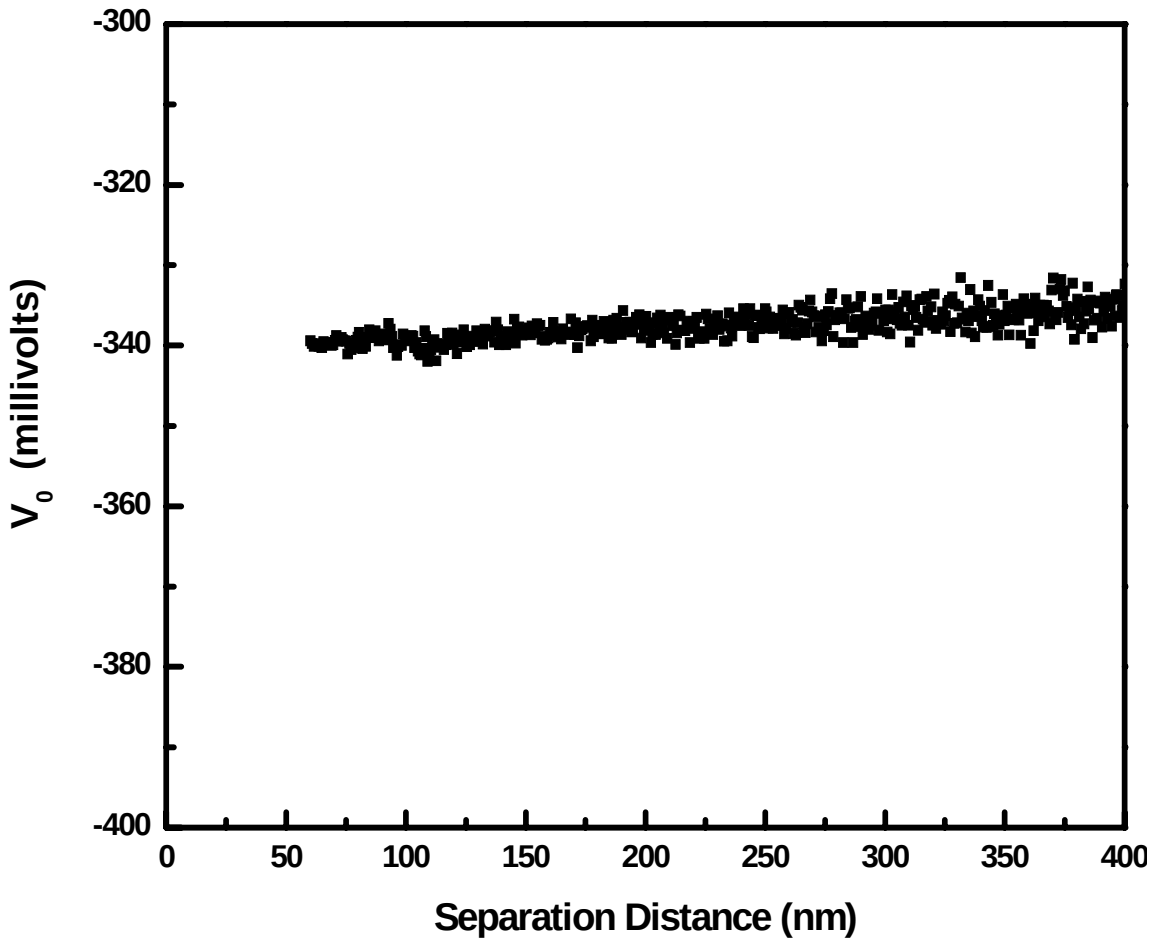

Figure 4: The residual sphere-plate potential difference shown as a function of the separation distance. The values correspond to the case of the high conductivity Si plate shown in figure 3(b) [22].

Such inhomogeneities can result from contamination which would lead to patches with different workfunctions. This requirement is particularly important as electrostatic forces resulting from such patches can mimic the distance dependence of the Casimir force [46]. Note that separation distances below 60 nm are always excluded due to the non-adiabatic response of the cantilever. At such small separations, the cantilever deflection is not stable and therefore cannot be used for inferring any experimental parameters or for measuring the Casimir force. Another important point to note is that in this procedure the measured $V_0$ is independent of the uncertainties in the calibration constant *km*, the separation distance *z*, and separation on contact $z_o$.

*5.3 Location of the sphere-plate contact*

The first step in the analysis process is the accurate determination of the point of sphere-plate contact, *C*. A typical measured cantilever deflection corresponding to an electrostatic force between the sphere and plate, as a function of the piezo extension *d*, is shown in figure 5 as squares. The region of sphere-plate contact, is encircled in figure 5 (after contact, the sphere is pushed up by the plate leading to an abrupt change in $S_{def}$). As shown in the blowup of this region in the inset, sphere-plate contact is achieved at a *d* somewhere between *X* and *Y*. Its exact value is sometimes unclear due to the finite data acquisition rate but can be obtained by extrapolation. The line *WX* from the two data points preceeding contact can be extended as shown by the dotted line to intersect the $S_{def}$ value corresponding to Y at *C* to obtain an accurate contact value $d_c$ by using: $d_X$ - ($d_{,W}$ - $d_X$) ($S_{def,X}$ - $S_{def,Y}$)/ ($S_{def,W}$ - $S_{def,X}$) , where the subscripts *W,X,Y* denote values at the respective points. Once *C* is determined, the $z_{piezo}$ can be obtained from *d* before contact as: $z_{piezo}= d_c - d$, where $d_c$ is the distance moved by the plate at the point of contact *C*.

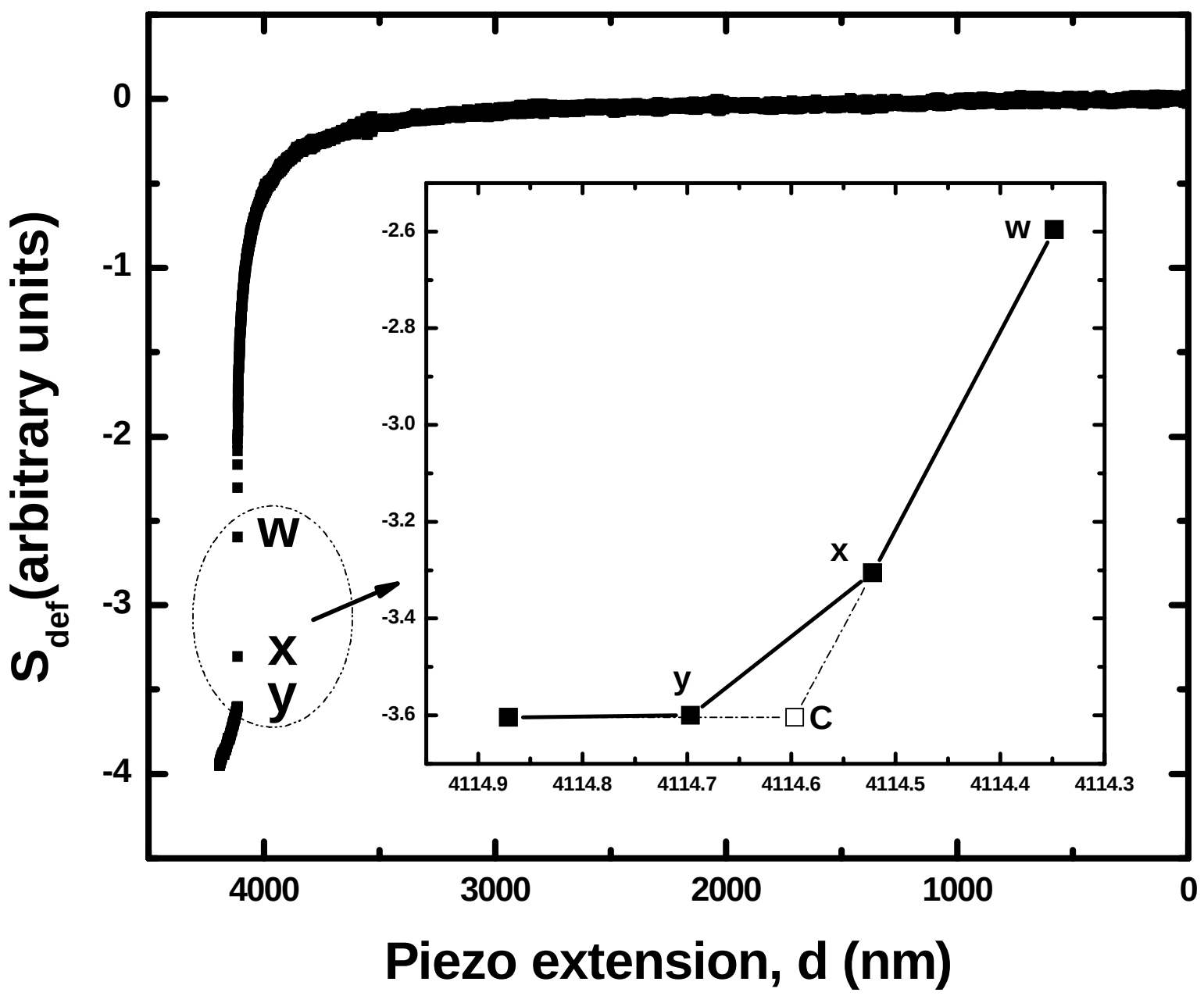

Figure 5: The measured $S_{def}$ corresponding to the electrostatic force due to an applied voltage between the sphere and plate as a function of the distance moved by the plate *d* is shown as squares. The region around the sphere-plate contact is shown in the inset. The slope of the line *WX* from the two data points preceding contact can be extrapolated to obtain an accurate value of the contact position *C* as explained in the text.

*5.4 Corrections for thermal and mechanical drift, and the measurement of the cantilever deflection per unit signal m*

The procedure discussed here is an advanced version of that provided by us in Ref.[33] and includes corrections for thermal noise of the cantilever and mechanical drift of the plate and the piezo. In figure 5, any linear change in $S_{def}$ at large sphere-plate separations of 5 μm indicates a systematic error due to scattered laser light, which should be subtracted. At these separations, the noise is far greater than the signal and in the absence of systematic errors the signal should average to zero and have no distance dependence. The deflection coefficient $m$ translates the cantilever deflection signal $S_{def}$ into the downward movement (in nm) of the cantilever tip in response to a force. In the electrostatic force measurement, larger applied voltage differences between the sphere and plate correspond to smaller values of $d_c$ and $S_{def}$ at sphere-plate contact. An example of this is shown in figure 6a where $d_c$ and $S_{def}$ at sphere-plate contact is plotted for the 28 different dc voltages applied to a high conductivity Si plate similar to that described in Ref. [22]. The time sequence of applied dc voltages is numbered from 1 through 28 and the time between each measurement is fixed. Note that due to thermal and mechanical drift, the points do not lie along a straight line but delineate an arrow head. At the base of the arrow head, the points are spread the widest due to having the largest time interval between the 1st and 28th measurements. At the tip of the arrow the measurements 14 and 15 overlap as the applied voltages are near $V_0$ and the time interval between them is small. This systematic error is corrected in the following manner. In the absence of any thermal noise or mechanical drift, point 28 (which corresponds to $S_{def}$ and applied sphere-plate difference voltages between points 1 and 2, see figure 3a) should lie at point A, along the line connecting points 1 and 2. This idea can be used to estimate the average drift between each measurement as: $[|(d_{c,1} - d_{c,2})*(S_{def,1} - S_{def,28})/(S_{def,1}-S_{def,2})|+(d_{c,1} - d_{c,28})]/27$, where $d_{c,i}$ are the contact points and $S_{def,i}$ are the corresponding deflections for the electrostatic measurement $i$. A second value of the average drift measurement is obtained from points 2,3 and 27(where $d_{c,27}$ is moved to point B). Note there are only 26 time intervals in this case. This process is repeated for five different set of points and the mean of the five average drift values is assumed to represent the mechanical and thermal drift of the system. This value of drift is used to correct the contact values of $d_{c,i}$ in figure 6a. For example, $d_{c,28}$ will have to be moved by 27 times this drift value. The corrected contact points are then plotted as a function of their corresponding $S_{def}$ in figure 6b. The best fit line to the data points gives the correct value of cantilever deflection coefficient $m$. Not applying the drift correction will lead to an error of order 0.4% for the case shown. Once the value of $m$ is determined all values of the separation distance in force measurements should be corrected for cantilever deflections as given in (1).

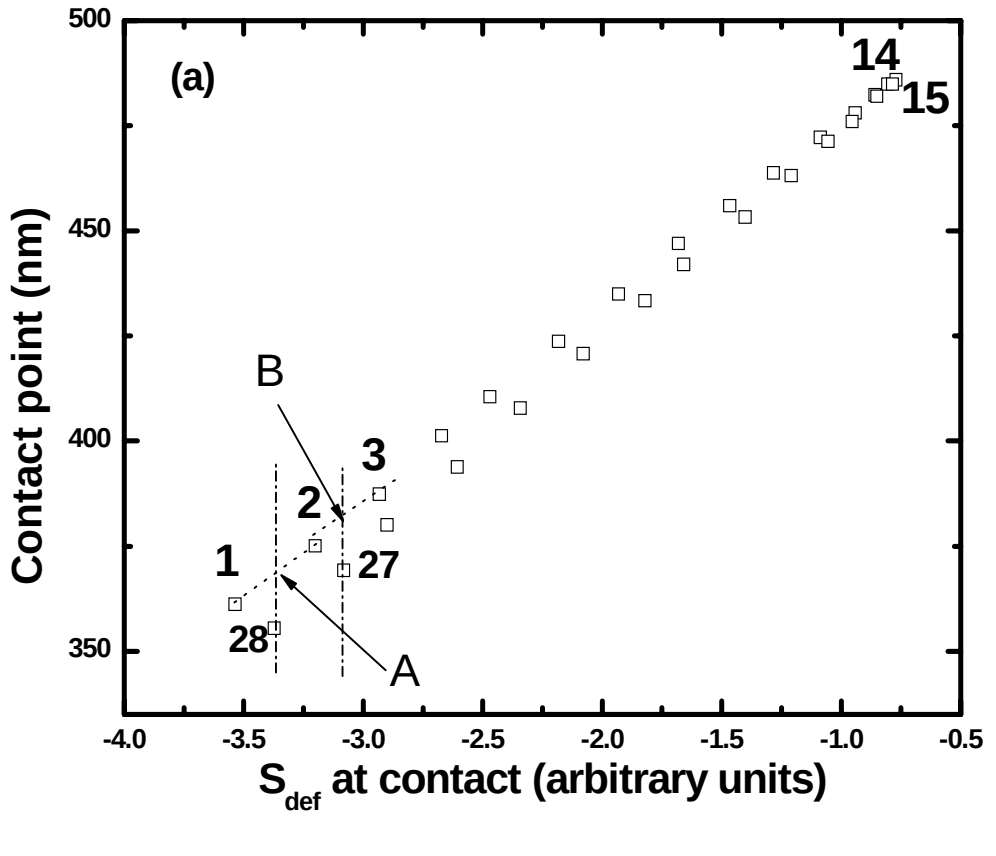

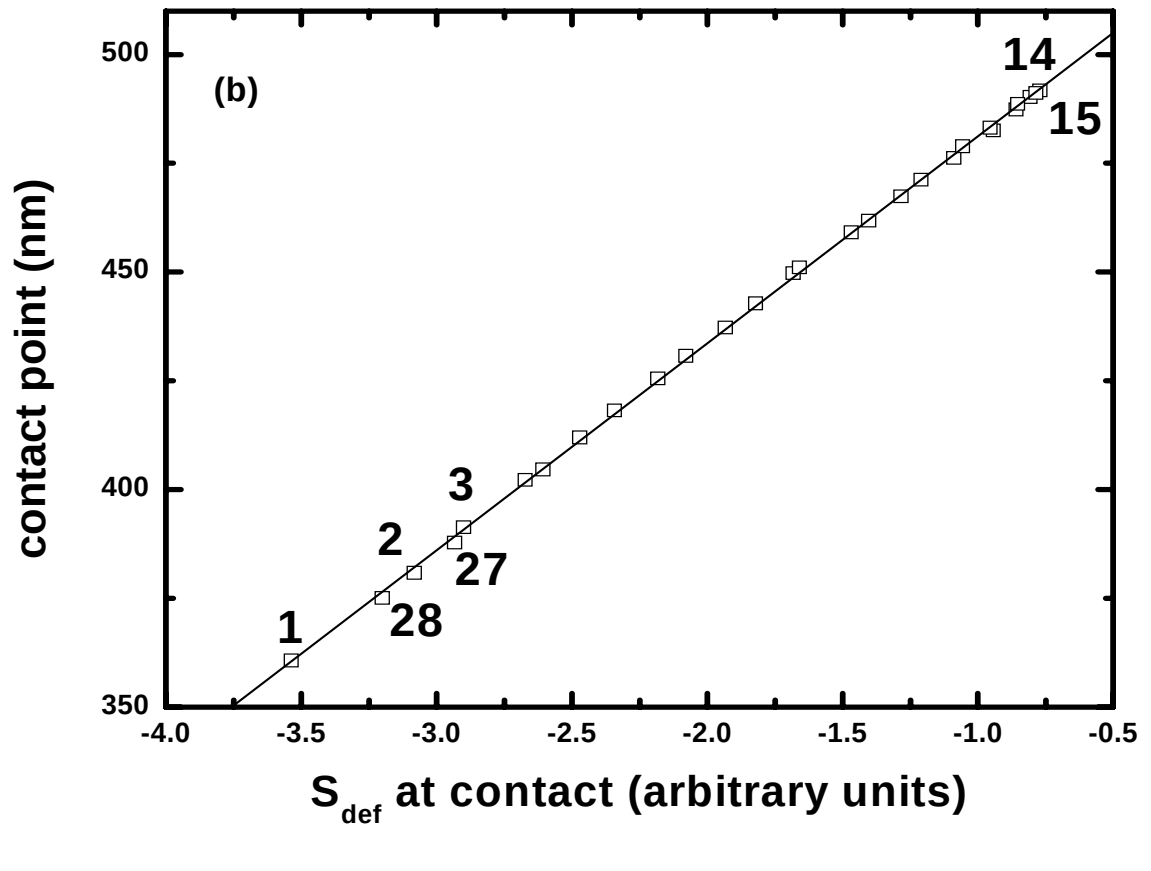

Figure 6: (a) The contact point $d_c$ and deflection signal $S_{def}$ at sphere-plate contact of the electrostatic force curves such as in figure 5. The numbers indicate the time sequence corresponding to the electrostatic force curves obtained by applying dc voltages to the plates. In (b) the contact points have been corrected for thermal and mechanical drift in the system.

*5.5 Measurement of the cantilever spring constant k and the average separation on contact $z_0$*

The value of *X(z)* (obtained from fitting the parabolic curves in figure 3) as a function of the sphere-plate separation *z*, can be used to obtain both the separation on contact $z_0$ and the cantilever spring constant *k*. Note in (3), that $k'=km$ can be treated as a single variable, where *m* has been independently determined in section 5.4. One can do a two unknown parameter least fit to *X(z)* find both values. However, in our experience an iterative fitting procedure, requiring the output value of only one unknown yields more robust and consistent results. We usually start with an approximate value for $z_o$ based on the highest roughness peaks on the sphere and plate, and determine the best fit value of $k'$, keeping $z_0$ fixed. In this fit the endpoint separation of the fit is kept at 2500 nm (larger endpoints have systematic errors due to perturbative expansion used in (4)). The first start separation for the fit is kept around 100 nm (smaller start points might have errors associated with space charge effects in the case of semiconductor plates). The best fit value for $k'$ is determined. The start separation is increased by 5 nm, the end separation kept fixed at 2500 nm and the fit is repeated to determine a second value of $k'$. This procedure is repeated for many other start separations between 100-300 nm. From these best fit values an average value of $k'$ is determined.

The average value of $k'$ obtained in the first iteration is used to fit the *X(z)* as a function of *z* to obtain a value for $z_0$. The same procedures with respect to the start and end point of the fit discussed above are used and an average value of $z_0$ from this first iteration is obtained. The iterative cycle is now repeated a second time with this value of $z_0$ and a new value of $k'$ is obtained. This value of $k'$ is used in the second iteration of finding a value of $z_0$. This procedure is repeated till $k'$ and $z_0$ (a) converge and (b) show only random variation as a function of the start point separation. The values of $k'$ and $z_0$ so determined are shown as a function of the start point of the *X(z)* fit in figures 7(a) and 7(b) respectively. Note the random variation $k'$ and $z_0$ on separation. Systematic errors, if present, would be indicated by a monotonic dependence on the start point separation used in the fit. The cantilever spring constant can be obtained from $k'$ with *m* obtained from section 5.3. In the case of $z_0$, the random errors can statistically be reduced far below 1 nm, by decreasing the spacing between start points used in the fit. However, to be conservative, the average roughness of the surfaces should be taken into account in setting the uncertainty.

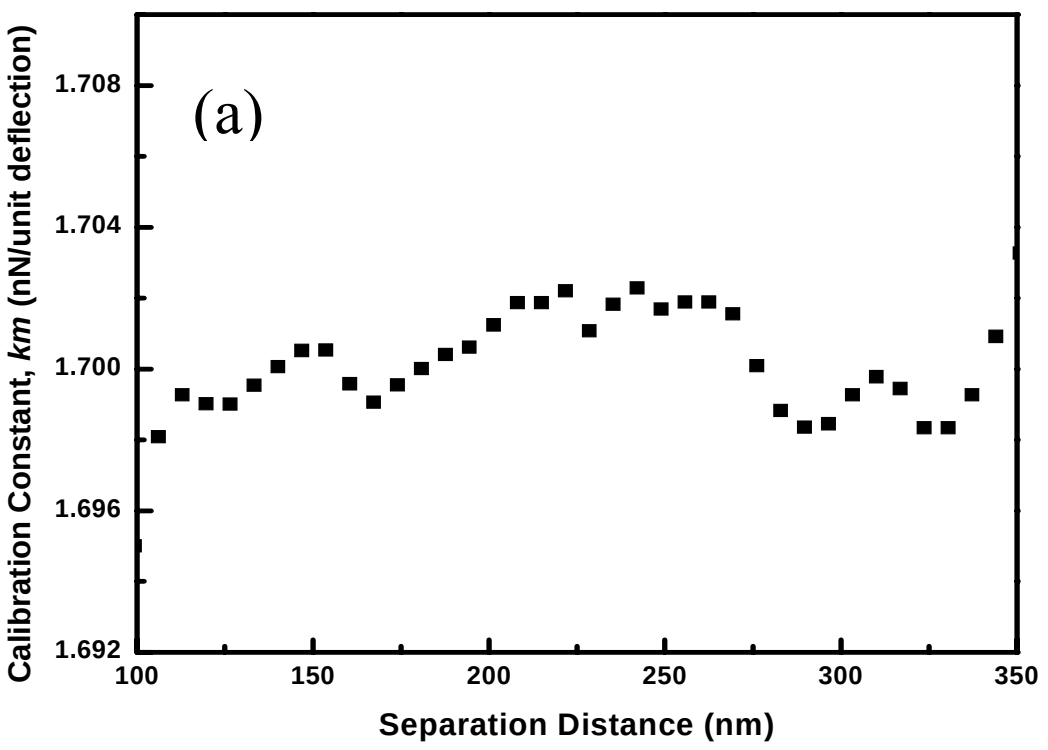

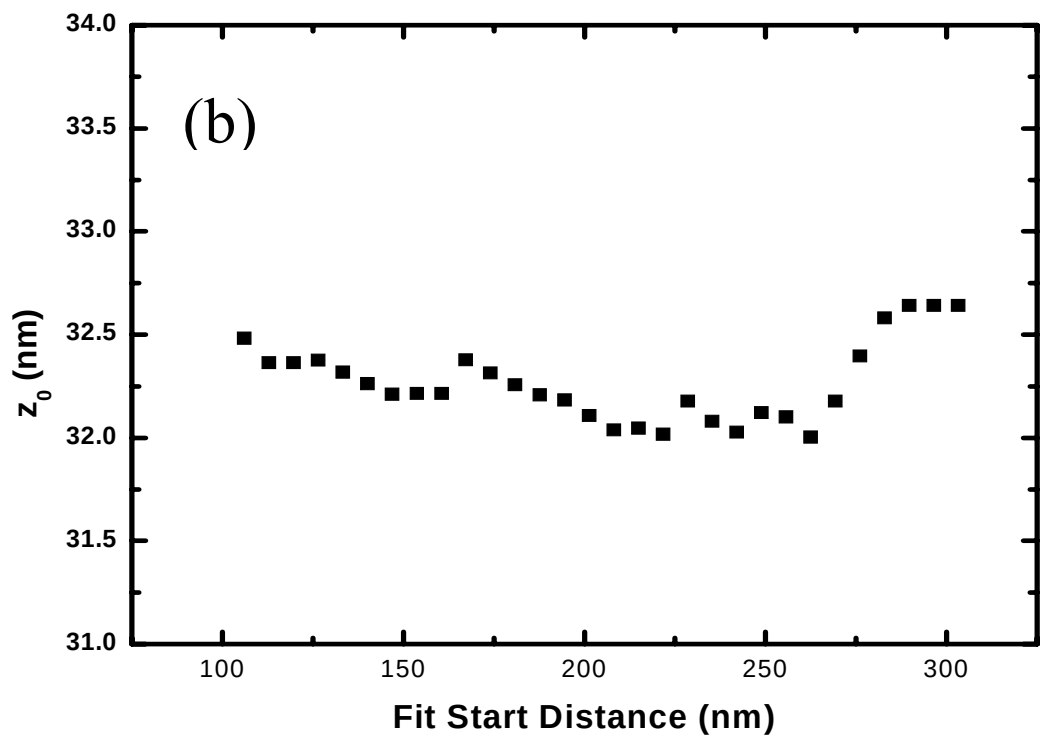

Figure 7: (a) The cantilever calibration constant *k'*, and (b) the average separation on sphere-plate contact $z_0$, obtained from fitting the *X(z)* as a function of the start point of the fit. Note the random variation of the values with the separation distance, which is indicative of the absence of systematic errors in the fit.

**6. Measurement of sphere and plate roughness**

In order to precisely compare the experimental measurement of the Casimir force with the theory, it is necessary to include the role of surface roughness. After measurement of the Casimir force, the topography of the bottom surface of the gold coated sphere and the plate used is measured with an AFM. The sphere is fixed on a glass plate surface and the topography of its bottom surface is scanned with the AFM. The roughness should be confirmed to have a stochastic distribution. The fractional surface area in each 1 nm bin is measured. As an example, table 1 lists the percent of the surface area corresponding to the various roughness heights for the case of Au sphere and Si plate used in Ref.[20]. A truncated table is shown here for illustration purposes only. The complete one is given in Ref. [20].

**Table 1.** Typical topography of the bottom of the Au coated sphere and Si plate as used in the precision measurement of the Casimir force in Ref. [20]

| **Au coated sphere** | | **Si plate** | |
|---|---|---|---|
| **Height (nm)** | **Percent Surface Area (%)** | **Height (nm)** | **Percent Surface Area (%)** |
| 0 | 0.008 | 0 | 0.002 |
| 1 | 0.085 | 0.1 | 0.081 |
| 2 | 0.12 | 0.2 | 0.884 |
| 3 | 0.16 | 0.3 | 4.27 |
| 4 | 0.41 | 0.4 | 10.384 |
| 5 | 0.48 | 0.5 | 34.379 |
| ….. | ….. | …. | …. |

The exact calculation of the roughness correction to the Casimir force using tables such as above has been discussed in previous publications [19-22]. Checks for correlated roughness corrections to the Casimir force [47] should be performed. In our case, these roughness corrections are much less than 1% and thus negligibly small. In general, we have found distortions with typical heights of 10-20 nm on the sphere and 0.3-0.6 nm on the Si plates. There are also rare (less than $10^{-5}$ of the surface area) needle like peaks on the gold coating of the sphere with heights up to 30 nm which are responsible for the final average separation on contact of the two surfaces. It is important to note that the height of the largest peaks observed in the topography should correspond to the average separation on contact measured in section 5 through the electrostatic force.

**7. Measurement of sphere diameter**

The diameter $2R$ of the gold coated sphere has to be independently measured for the calculation of the theoretical force. It is also required for the determination of the experimental parameters such as the cantilever spring constant $k$ and the average separation on contact $z_0$ using (4). The sphere diameter is measured at the end of the experiment after the completion of the Casimir force measurement. A scanning electron microscope (SEM) is used in the measurement. A special multi step calibration of the SEM has to be done to achieve the necessary precision of ±0.03 μm. A gold or platinum coated silicon AFM calibration grid of 1 μm pitch and <0.5 μm thick walls is used as the standard. However, this AFM grid itself has to be calibrated. The first step in this is the calibration of the lateral movement of the AFM piezos. The same optical interferometer [43] used in section 4 is used to calibrate the AFM piezo. This calibrated movement of the piezo is next used to calibrate the grid using its AFM topographic images. Now the independently calibrated grid is ready to be used as an SEM calibration standard for measuring the sphere diameter.

## 8. Measurement of normal Casimir force

The normal Casimir force between the gold coated sphere and plate is best done after the electrostatic force measurements. Note that while $V_0$ values between similar metals are small, when different materials such as the Au sphere and a semiconductor plate are used, $V_0$ values between 200-400 mV are possible due to the differences in the work functions. Very large values of the residual potential difference $V_0$ and/or contact separation $z_0$ which varies with sphere-plate separation are indicative of the presence of dust or other impurities on the sphere or plate surface. Samples exhibiting such properties should not be used in precision Casimir force measurments, as the systematic errors from the residual electrostatic force cannot be consistently subtracted.

We have provided extensive discussions of the exact procedure for the normal Casimir force between gold spheres and plates in many recent publications [19-22]. The Casimir force is measured as a function of the separation distance, after compensating the electrostatic force by applying a voltage $V_0$ to the plate while the sphere remains grounded. Alternatively, a value for the Casimir force at each separation distance can be obtained from the electrostatic force measurements in section 5. For example, in figure 3, the residual deflection from the top of the parabola corresponds to the Casimir force. In precision measurements of the Casimir force, we have found that both methods lead to overlapping results within the experimental error. This coincidence between the two methods is an important check of the validity of the results.

## 9. Conclusions

We have discussed the methods, measurements and the related data analysis involved in the determination of the critical experimental parameters required for precision tests of the normal Casimir force using an AFM. The need for an oil-free high vacuum system in both the experimental chamber and the thin film metal evaporator was emphasized. The fabrication of a thermally stable and vacuum compatible sphere-cantilever system and its appropriate coating with an Al and Au layer was presented. Steps to correct the contact point between the two surfaces for thermal and mechanical drift were explained. The importance of procedures to obtain consistent and separation independent values for the average separation on contact $z_0$, the residual potential difference $V_0$, and the cantilever spring constant $k$ were discussed in detail. Given the stringent requirements of measuring and subtracting residual electrostatic force only metallic or highly conductive materials are appropriate for precision measurements of the Casimir force and dielectric surfaces cannot be used. In this regard even a metal layer with organic coating as used in Ref. [48] is inappropriate. The measurement procedures outlined should result in Casimir force measurements using the AFM with random and systematic errors of 1% as is rigorously shown in [21,40].

## 10. Acknowledgements

The experimental section of this work was supported by NSF Grant No.PHY0653657 and the instrumental supplies along with the analysis by DOE Grant No. DE-FG02-04ER46131. The authors acknowledge help from G.L. Klimchitskaya and V.M. Mostepanenko with the theoretical analysis.